\documentclass[12pt]{article}
\usepackage{graphicx}
\usepackage{endfloat}
\begin{document}
\title{ Experimental study of quantum decoherence using nuclear magnetic resonance}
\author{\small{} Jingfu Zhang$^{1}$ , Zhiheng Lu$^{1}$,Lu Shan$^{2}$,and Zhiwei
Deng$^{2}$  \\
\small{} $^{1}$Department of Physics,\\
\small{}Beijing Normal University, Beijing,
100875, Peoples' Republic of China\\
\small{} $^{2}$Testing and Analytical Center,\\
\small{}  Beijing Normal University,
 Beijing,100875, Peoples' Republic of China}
\date{}
\maketitle
\begin{center}\bf Abstract\end{center}

Quantum decoherence has been studied using nuclear magnetic
resonance(NMR). By choosing one qubit to simulate environment, we
examine the decoherence behavior of two quantum systems: a one
qubit system and a two qubit system. The experimental results show
 agreements with the theoretical predictions. Our experiment
schemes can be generalized to the case that the environment is
composed of multiple qubits.
\\
PACS number(s):03.67
\begin{center}\bf 1. INTRODUCTION\end{center}

Quantum decoherence is a purely quantum-mechanical effect through
which a system loses its coherence behavior by getting entangled
with its environment degrees of freedom [1][2]. A real quantum
system always interacts with its surrounding environment. The time
evolution induced by the interaction introduces entanglement
between the system and  environment when the system initially lies
in a superposition of states. Quantum decoherence can be viewed as
the consequence of such entanglement. When the state of the system
is described by the reduced density matrix through tracing over
the environment degrees of freedom, decoherence makes the
off-diagonal matrix elements approach 0, and leaves diagonal ones
unaltered. Quantum decoherence is thought as a main obstacle for
experimental implementations of quantum computation and has been
widely studied in theory [3]-[5]. Some theoretical and
experimental schemes have also been proposed to solve the problem
of decoherence in order to fully use coherence in quantum
information[6]-[8]. Myat.C.J. et al observed and studied
decoherence by coupling an ion in a Paul trap to a reservoir that
can be controlled. Their experimental results tested their
theoretical predictions[9][10].

  As a useful tool to research the microscopic world through
the macroscopic signal, nuclear magnetic resonance can be used to
study the fundamental problems in quantum mechanics. Besides some
quantum algorithms have been implemented on NMR quantum computers,
a quantum harmonic oscillator was simulated by S. Somaroo et al
[11], and the complementarity was tested by X.Zhu et al [12]. In
this paper, we will study quantum decoherence using NMR. Although
we choose one nucleus to simulate the environment, our schemes can
be generalized to the case that the environment is composed of
multiple nuclei.

\begin{center}\bf 2. The decoherence behavior of a one qubit system\end{center}

For the simplest case, we choose a sample containing two coupled
spin 1/2 nuclei to implement experiments. The two nuclei are
defined as qubit 1 and qubit 2 which represent the quantum system
and environment, respectively. The experimental scheme is shown in
Fig.1. $R(\theta)$ denotes a rotation operation for qubit 2, and
$H$ denotes the Walsh-hadamard transform for qubit 1 [13].
Experiments start with the pseudo-pure initial state
$|\downarrow>_{1}|\downarrow>_{2}$ by using spatial
averaging[14][15], where $|\downarrow>_{i}$ (i=1 or 2) denotes the
spin 'down' state of qubit $i$. $H$ and $R(\theta)$ transform
$|\downarrow>_{1}|\downarrow>_{2}$ into
$(|\uparrow>_{1}-|\downarrow>_{1})(a(\theta)|\uparrow>_{2}+b(\theta)|\downarrow>_{2})/\sqrt{2}$,
where $|a|^{2}+|b|^{2}=1$. After the control-not operation is
applied, the system and environment lie in the entangled state
described as
\begin{equation}\label{1}
  \rho_{s+e}=\left(\begin{array}{cccc}
    |a|^{2} & ab^{*} & -ab^{*} & -|a|^{2} \\
    a^{*}b & |b|^{2} & -|b|^{2} & -a^{*}b \\
    -a^{*}b & -|b|^{2} & |b|^{2}& a^{*}b \\
    -|a|^{2} & -ab^{*} & ab^{*} & |a|^{2} \
  \end{array}\right),
\end{equation}
where a constant factor is ignored. The state of the system is
represented by the reduced density matrix[16]

\begin{equation}\label{2}
 \rho_{s}=\left(\begin{array}{cc}
   1 & -ab^{*}-a^{*}b \\
   -a^{*}b-ab^{*} & 1 \
 \end{array}\right)
\end{equation}
where the off-diagonal elements describe the coherence behavior.
When $\theta$ is chosen as different values, entangled states with
different amount of entanglement are gotten [17]. It is clear that
the coherence of the system depends on $\theta$.

 Our experiments use carbon-13 labelled chloroform
dissolved in d6-acetone and carbon-13 trichloroethylene(TCE)
dissolved in d-chloroform as two samples. TCE's structure is shown
in Fig.2. $^{1}$H is decoupled during the experiments. The two
$^{13}$C nuclei are assigned as qubit 1 and qubit 2, respectively.
 Data are taken at controlled temperature (22$^{0}C$) with the Bruker DRX 500 MHz
 spectrometer of Beijing Normal University. For chloroform,
 the resonance frequencies
 $\nu_{1}=125.76$ MHz for $^{13}$C,
 and $\nu_{2}=500.13$ MHz for $^{1}H$.
 The coupling constant $J_{12}$ is measured to be 215 Hz. For TCE,
 the resonance frequencies  $\nu_{1}=125.7869009$MHz (117ppm in the spectrum),
 $\nu_{2}=\nu_{1}+903.6Hz$. The coupling constant $J_{12}=103.1$ Hz.

 The pulse sequence
  $[\theta]_{x}^{2}-[\frac{\pi}{2}]_{x}^{1,2}-
  \frac{1}{4J_{12}}-[\pi]_{x}^{1,2}-\frac{1}{4J_{12}}-[\frac{\pi}{2}]_{y}^{2}$
 transforms the initial state $|\downarrow>_{1}|\downarrow>_{2}$
 into an entangled state. $[\theta]_{x}^{2}$ denotes a rectangular pulse along
 x-axis for qubit 2, and $\frac{1}{4J_{12}}$ denotes the
evolution caused by the magnetic field for $\frac{1}{4J_{12}}$
when pulses are closed. In NMR experiments, the entangled state
can be described by the following deviation density matrix[18]

\begin{equation}\label{3}
  \rho_{pp}=\left(\begin{array}{cccc}
    cos^{2}(\frac{\theta}{2}) & sin(\frac{\theta}{2})cos(\frac{\theta}{2}) & -sin(\frac{\theta}{2})cos(\frac{\theta}{2}) & -cos^{2}(\frac{\theta}{2}) \\
    sin(\frac{\theta}{2})cos(\frac{\theta}{2}) & sin^{2}(\frac{\theta}{2}) & -sin^{2}(\frac{\theta}{2}) & -sin(\frac{\theta}{2})cos(\frac{\theta}{2}) \\
    -sin(\frac{\theta}{2})cos(\frac{\theta}{2}) & -sin^{2}(\frac{\theta}{2}) & sin^{2}(\frac{\theta}{2})& sin(\frac{\theta}{2})cos(\frac{\theta}{2}) \\
    -cos^{2}(\frac{\theta}{2}) & -sin(\frac{\theta}{2})cos(\frac{\theta}{2}) & sin(\frac{\theta}{2})cos(\frac{\theta}{2}) &  cos^{2}(\frac{\theta}{2})\
  \end{array}\right).
\end{equation}
This is the case that $a=cos(\frac{\theta}{2})$, and
$b=sin(\frac{\theta}{2})$ in Equation(1). By tracing $\rho_{pp}$
over the environment degrees of freedom, the state of the system
is described as

\begin{equation}\label{4}
 \rho_{spp}=\left(\begin{array}{cc}
   1 & -sin\theta \\
    -sin\theta & 1 \
 \end{array}\right),
\end{equation}
where the dependence of the coherence on $\theta$ can be
represented by a sine function. In the NMR spectrum, the sum of
the amplitudes of the two peaks of $^{13}$C is proportional to the
off-diagonal element in equation (4).

 Fig.3a and Fig.3b show the experimental results when the sample is
 chosen as chloroform. When $^{13}$C and $^{1}$H lie in an entangled state,
 the amplitudes of the two NMR peaks of $^{13}$C
 are almost equal as shown in Fig.3a, where $\theta=50.3^{\circ}$.
 When $\theta$ is chosen as different values, the dependence of
 the coherence on $\theta$ is shown in Fig.3b. The experimental
 data points can be fitted as a sine-shaped
 curve with the period $1.96\pi$. The error is about $2.0\%$. It mainly results
 from the imperfection of pulses. When $\theta=0$ or $\pi$, $^{13}$C and $^{1}$H
 lie in a maximally entangled state [18]. If we ask only about the
 state of $^{13}$C, it lies in a completely mixed-state, in
which the coherence completely disappears[16]. When $\theta=\pi/2$
or $3\pi/2$ , $^{13}$C and $^{1}$H lie in a product state. The
coherence of $^{13}$C is maximal. We will discuss this case in
detail. If the pulse sequence
$[\theta]_{x}^{2}-[\frac{\pi}{2}]_{x}^{1,2}$ is applied to the
initial state $|\downarrow>_{1}|\downarrow>_{2}$, $^{13}$C and
$^{1}$H are still in a product state. The spectrum of $^{13}$C is
shown in Fig.4a, where $\theta=50.3^{\circ}$. The amplitudes of
two peaks varies versus $\theta$ shown in Fig.4b, in which they
are marked by '$\triangle$' and '$\bigcirc$',respectively.
Nevertheless, their sum, which is corresponding to the coherence
of the system, is independent of $\theta$ as shown by the line
marled by '*' in Fig.4b.

Experimental results of TCE are shown by Fig.5a and Fig.5b. A
carbon spectrum is shown in Fig.5a when $\theta=27^{\circ}$. The
dependence of the coherence on $\theta$ is shown in Fig.5b. The
period of the sine-shape curve is $1.83\pi$. The error is about
$8.5\%$. Because the difference of the two $^{13}$C nuclei in TCE
is much less than the difference of $^{13}$C and $^{1}$H in
chloroform, the selective pulses used in the latter experiment are
less perfect than the former experiment. The error is larger when
the sample is changed to TCE.

 Our scheme can be generalized to the case
 that the environment is composed of
$N(N>1)$ qubits. The quantum network is shown in Fig.6. For
convenience, we set the rotation operation as the same form
$R(\theta)$. After the system gets entangled with qubits in
environment one by one, its state is represented as
\begin{equation}\label{5}
 \rho_{spp}=\left(\begin{array}{cc}
   1 & -sin^{N}\theta \\
    -sin^{N}\theta & 1 \
 \end{array}\right).
\end{equation}
When $N\rightarrow \infty$, the coherence approach 0, if
$sin\theta\neq\pm 1$. In our generalized scheme, the qubits in
environments are independent. We call this mode "ideal gas" mode.
The result above is in agreement with the one which is gotten by
harmonic oscillator mode in theory [3].

\begin{center}\bf 3.The  decoherence behavior of a two
qubit system in an entangled state\end{center}

 If a two qubit system lies in an entangled state, the off-diagonal
elements in its density matrix also represent the coherence. We
choose carbon 13-labelled trichloroethylene dissolved in
d-chloroform as a sample. The two $^{13}$C nuclei, which are
defined as qubit 1 and qubit 2, are assigned as the system, and
the $^{1}$H, which is defined as qubit 3, is assigned as
environment. The influence of the other nuclei can be ignored. The
pseudo-pure initial state $|\downarrow>_{1}|\downarrow>_{2}$ is
also prepared by spatial averaging [11]. $^{1}$H is decoupled
until the entangled state of the system is prepared. We choose
composite pulse decoupling, and NOE can be ignored due to the
short decoupling time. The following pulse sequence
$[\frac{\pi}{2}]_{x}^{1,2}-
\frac{1}{4J_{12}}-[\pi]_{x}^{1,2}-\frac{1}{4J_{12}}-[\frac{\pi}{2}]_{y}^{2}$
 transforms $|\downarrow>_{1}|\downarrow>_{2}$ into a basis entangled
 state [18]
\begin{equation}\label{6}
  \rho_{s}(0)=I_{x}^{1}I_{x}^{2}-I_{z}^{1}I_{z}^{2}-I_{y}^{1}I_{y}^{2},
\end{equation}
where $I_{x}^{k}$, $I_{y}^{k}$ and $I_{z}^{k}(k=1,2)$ are
respectively the matrices for the three components of the angular
momentum of the spins. As soon as the system lies in the entangled
state, the decoupling pulses are closed. The system and
environment will evolute in the magnetic field. In the rotating
frame, the Hamitonian is represented as
\begin{equation}\label{7}
  H=\Delta\omega_{12}I_{z}^{1}+\Delta\omega_{32}I_{z}^{3}+2\pi
  J_{12}I_{z}^{1}I_{z}^{2}+2\pi J_{23}I_{z}^{2}I_{z}^{3}+2\pi
  J_{13}I_{z}^{1}I_{z}^{3}
\end{equation}
 where $\omega_{i}/2\pi(i=1,2,3)$ are the
resonance frequencies of spins 1,2, and 3,
 $\Delta\omega_{12}=\omega_{1}-\omega_{2}$,
 $\Delta\omega_{32}=\omega_{3}-\omega_{2}$, and $\hbar$ is set to 1.
 We choose $\omega_{2}/2\pi$ as the frequency of the rotating frame.
By applying a hard (nonselective) pulse $[\pi]_{x}^{1,2,3}$ in the
middle of an evolution period t, the chemical shift evolution
during this period can be refocused [20][21], so that the system
in the entangled state evolutes under
\begin{equation}\label{8}
  H_{eff}=2\pi J_{12}I_{z}^{1}I_{z}^{2}+2\pi J_{23}I_{z}^{2}I_{z}^{3}+2\pi
  J_{13}I_{z}^{1}I_{z}^{3},
\end{equation}
where the last two terms describe the interaction between the
system and environment. Only these two terms cause the system to
evolute, because entangled basis states are the eigenstates of
$I_{z}^{1}I_{z}^{2}$ [19]. After evolution time t, by tracing over
the environment degrees of freedom, the state of the system can be
represented as
\begin{equation}\label{9}
  \rho_{s}(t)=\left(\begin{array}{cccc}
    -\frac{1}{2} & 0 & 0 & cos(\varphi_{13}+\varphi_{23}) \\
    0 & \frac{1}{2} & 0 & 0 \\
    0 & 0 & \frac{1}{2}& 0 \\
    cos(\varphi_{13}+\varphi_{23}) & 0 & 0 & -\frac{1}{2} \
  \end{array}\right),
\end{equation}
where $\varphi_{13}=\pi J_{13}t$, and $\varphi_{23}=\pi J_{23}t$.
In NMR experiments, this equation is equivalent to the following
one
\begin{equation}\label{10}
  \rho_{s}(t)=\left(\begin{array}{cccc}
    1 & 0 & 0 & -cos(\varphi_{13}+\varphi_{23}) \\
    0 & 0 & 0 & 0 \\
    0 & 0 & 0& 0 \\
    -cos(\varphi_{13}+\varphi_{23}) & 0 & 0 & 1 \
  \end{array}\right)
\end{equation}
The off-diagonal elements are called double quantum coherence,
which can be observed through a readout pulse
$[\frac{\pi}{2}]_{x}^{2}$ which transforms the system into the
state represented as
\begin{equation}\label{11}
 \rho_{sr}(t)=\left(\begin{array}{cccc}
    0 & i & -icos(\varphi_{13}+\varphi_{23}) & cos(\varphi_{13}+\varphi_{23}) \\
    -i & 0 & cos(\varphi_{13}+\varphi_{23}) & icos(\varphi_{13}+\varphi_{23}) \\
    icos(\varphi_{13}+\varphi_{23}) & cos(\varphi_{13}+\varphi_{23}) & 0& -i \\
    cos(\varphi_{13}+\varphi_{23}) & -icos(\varphi_{13}+\varphi_{23}) & i & 0 \
  \end{array}\right).
\end{equation}
The coherence behavior can be observed in the NMR spectrum.

Experimental data are also taken at controlled temperature
(22$^{0}C$) with a Bruker DRX 500 MHz spectrometer. The coupling
constants $J_{12}=103.1$Hz, $J_{23}=201.3$Hz, and $J_{13}=9.23$Hz.
$^{1}$H is again decoupled during recording the FID signal in
order to simplify the spectrum. From the view of quantum
mechanics, decoupling also can be thought as the process of
tracing over the environment degrees of freedom to get the reduced
density matrix of the system. We will discuss this point in detail
in another paper. Fig.7a is the carbon spectrum through the
readout pulse $[\frac{\pi}{2}]_{x}^{2}$ when the evolution time is
3.50ms. Fig.7b shows that the amplitude of the left peak of C1(see
Fig.2) in the spectrum varies versus the evolution time. The
experimental data points can be fitted as function $5.8cos(2\pi
 t/T)$, where $T=8.72ms$. In theory, $T = 9.50$ms. The error
 is about $8.2\%$. It mainly results from the imperfection
of pulses, and the effect of decoherence which cannot be
controlled.

If environment is composed of multiple qubits, the interaction
between the system and environment is represented as
\begin{equation}\label{12}
  H_{i}=\sum_{k=3}^{N+2}(2\pi J_{1k}I_{z}^{1}I_{z}^{k}+2\pi
  J_{2k}I_{z}^{2}I_{z}^{k}).
\end{equation}
Under the evolution induced by $H_{i}$, the state of the system is
described as
\begin{equation}\label{13}
  \rho_{12}(t)=I_{x}^{1}I_{x}^{2}\prod_{k=3}^{N+2}cos(\varphi_{1k}+\varphi_{2k})
  -I_{z}^{1}I_{z}^{2}-I_{y}^{1}I_{y}^{2}\prod_{k=3}^{N+2}cos(\varphi_{1k}+\varphi_{2k}),
\end{equation}
where $\varphi_{1k}=\pi J_{1k}t$, and $\varphi_{2k}=\pi J_{2k}t$.
If $ N\rightarrow \infty$, the coherence usually approachs to 0
very fast,if $t\neq 0$. The decoherence of the system is also
caused by the entanglement between the system and environment.

\begin{center}\bf 4. Conclusion \end{center}

In our experiments, we examine quantum decoherence of two systems
by two experimental schemes. For the one qubit system, the
entanglement between the system and environment is realized a
quantum network. For the two qubit system, such entanglement
results from the evolution induced by the interaction between the
system and environment. Although the two schemes seem different,
the interaction is their common prerequisite. We can get the
results when the environment is composed of multiple qubits by
generalizing our schemes. It is unnecessary to consider the state
of the environment. In fact, we usually cannot describe the state
of environment clearly. Through the entanglement between the
system and environment, the environment degrees of freedom are
introduced into the state of the system. Decoherence is the result
of tracing over such degrees of freedom. If real environment
consists of multiple particles each of which has more than two
degrees of freedom, our results are still correct.

\begin{center}\bf Acknowledgements \end{center}

This work was partly supported by the National Nature Science
Foundation of China. We are also grateful to Professor Shouyong
Pei of Beijing Normal University for his helpful discussions on
the theories in quantum mechanics and Miss Jinna Pan and for her
help during experiments.
\newpage
\bibliographystyle{article}

\newpage
{\begin{center}\large{Figure Captions}\end{center}
\begin{enumerate}
\item Quantum network used to study the decoherence
behavior of a one qubit system. Qubit 1 and qubit 2 are defined as
the system and environment, respectively. $H$ denotes the
Walsh-hadamard transform, and $R(\theta)$ is a rotation for qubit
2.
\item The structure of trichloroethylene. The two $^{13}C$ nuclei are
assigned as qubit 1 and qubit 2, respectively.
\item The experimental results when $^{13}$C and $^{1}$H of chloroform
lie in entangled states. Fig.3a is the carbon spectrum when
$\theta=50.3^{\circ}$. Fig.3b shows the dependence of
 the coherence on $\theta$. The experimental data points can be
 fitted as a sine-shaped curve with the period $1.96\pi$.
\item The experimental results when $^{13}$C and $^{1}$H of chloroform
lie in product states. Fig.4a is the carbon spectrum when
$\theta=50.3^{\circ}$. The amplitudes of two peaks varies versus
$\theta$ are also shown in Fig.4b, in which they are marked by
'$\triangle$' and '$\bigcirc$', respectively. Their sum, which is
corresponding to the coherence of the system, is independent of
$\theta$ as shown by the line marked by '*'.
\item The experimental results when two $^{13}$C
nuclei of trichloroethylene lie in entangled states. Fig.5a is the
carbon spectrum when
 $\theta=27^{\circ}$. Fig.5b shows the dependence of
 the coherence on $\theta$. The experimental data points can be
 fitted as a sine-shaped curve with the period
 $1.83\pi$. $^{1}$H is decoupled during the whole experiments.
\item The generalized experimental scheme when environment
is composed of qubit 2, qubit 3, ..., and qubit N+1. Qubit 1
denotes the system.
\item The experimental results when the two $^{13}$C nuclei of
trichloroethylene in the entangled state evolute under
$H_{eff}$(see the text). Fig.7a is the carbon spectrum through the
readout pulse $[\frac{\pi}{2}]_{x}^{2}$ when the evolution time is
3.50ms. Fig.7b shows that the amplitude of the left peak of
C1(shown in Fig.2) varies versus the evolution time which is
denoted as t. The experimental data points can be fitted as
function $5.8cos(2\pi t /T)$, where $T=8.72ms$. $^{1}$H is
decoupled during recording the FID signal.
\end{enumerate}
\begin{figure}{1}
\includegraphics[]{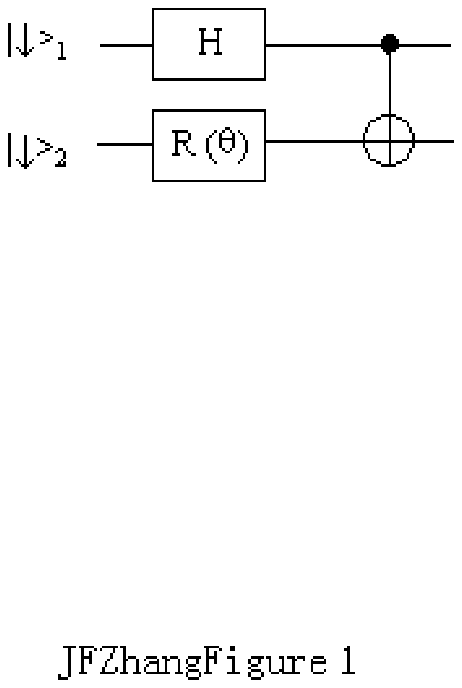}
\caption{}
\end{figure}
\begin{figure}{2}
\includegraphics[]{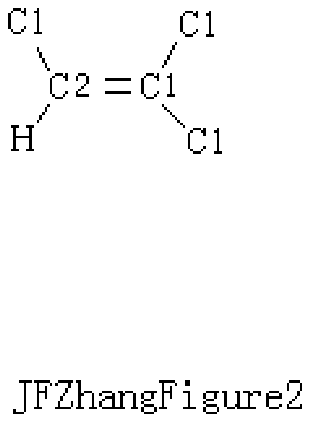}
\caption{}
\end{figure}
\begin{figure}{3}
\includegraphics[]{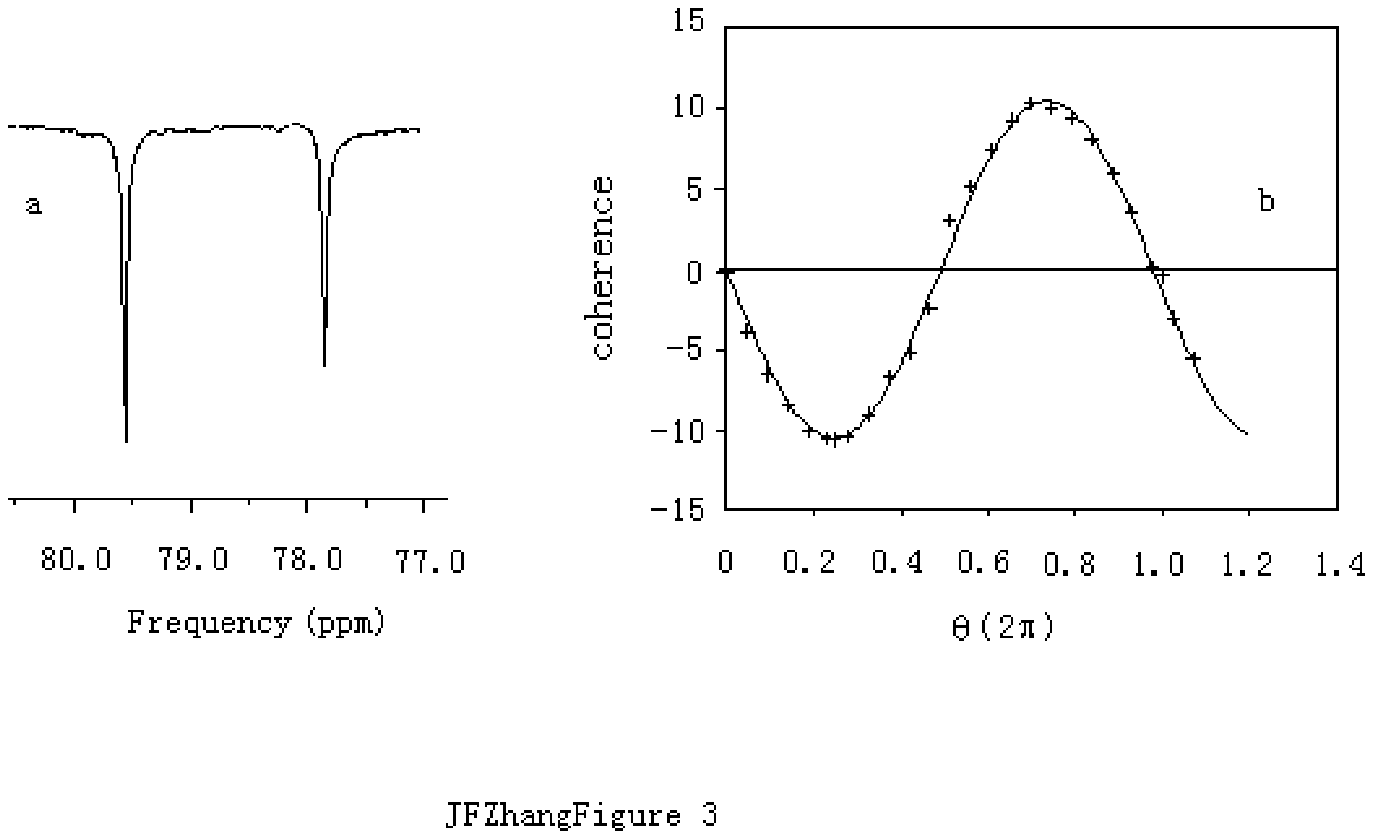}
\caption{}
\end{figure}
\begin{figure}{4}
\includegraphics[]{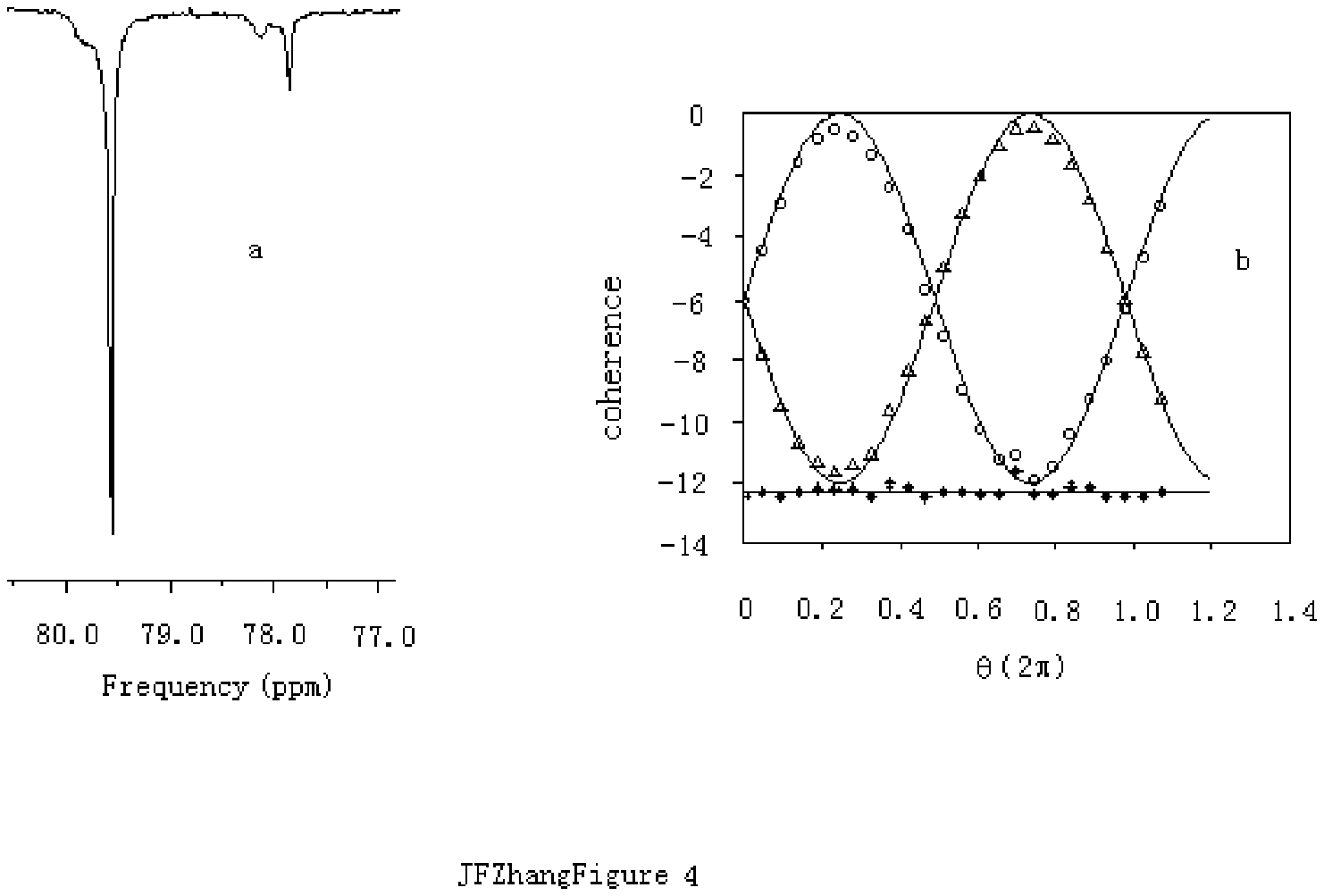}
\caption{}
\end{figure}
\begin{figure}{5}
\includegraphics[]{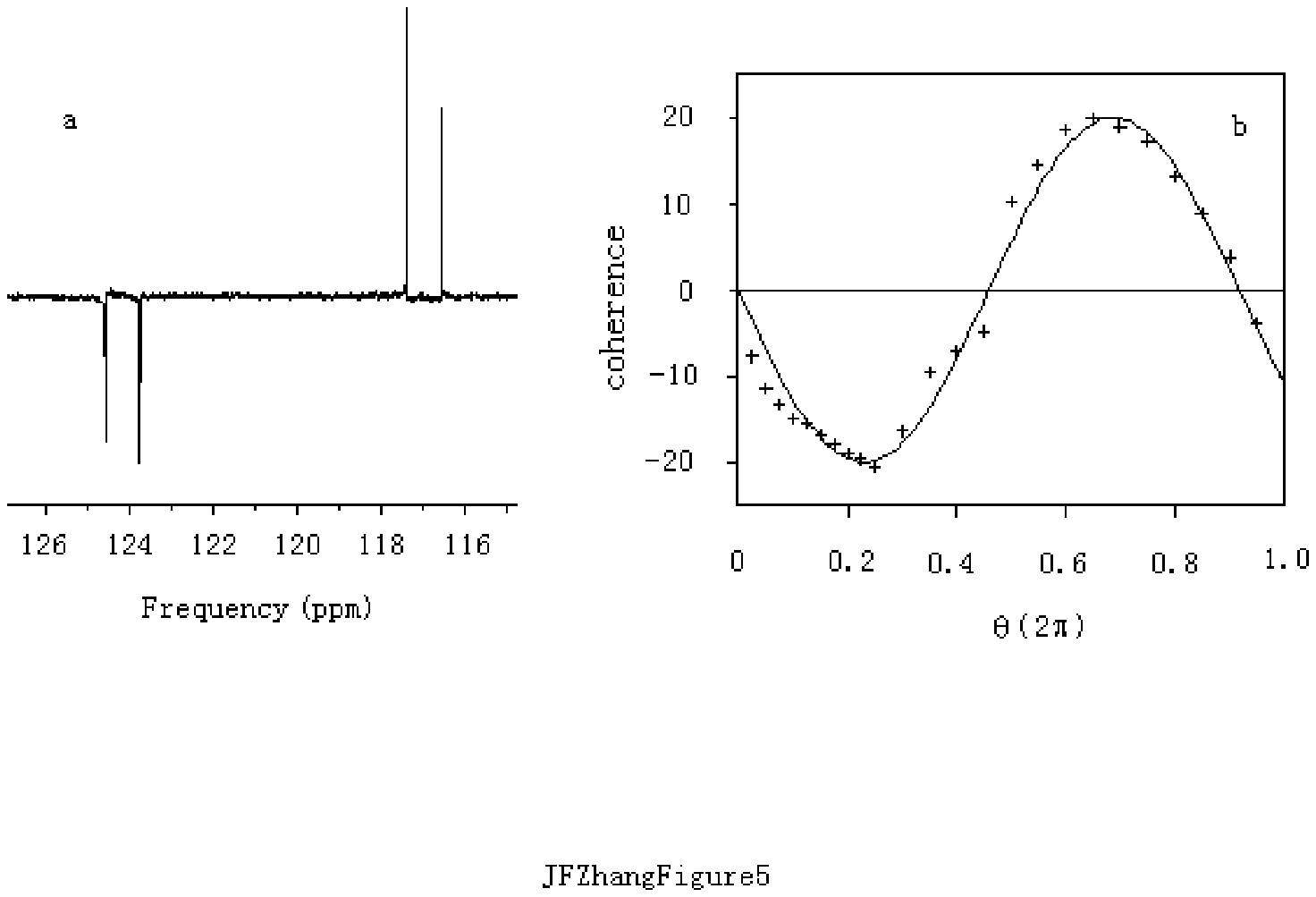}
\caption{}
\end{figure}
\begin{figure}{6}
\includegraphics[]{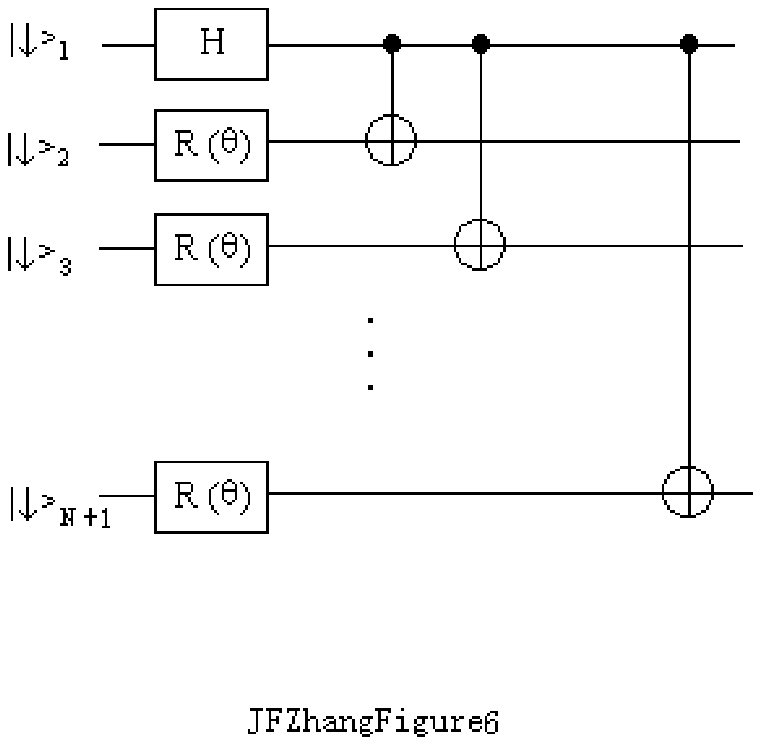}
\caption{}
\end{figure}
\begin{figure}{7}
\includegraphics[]{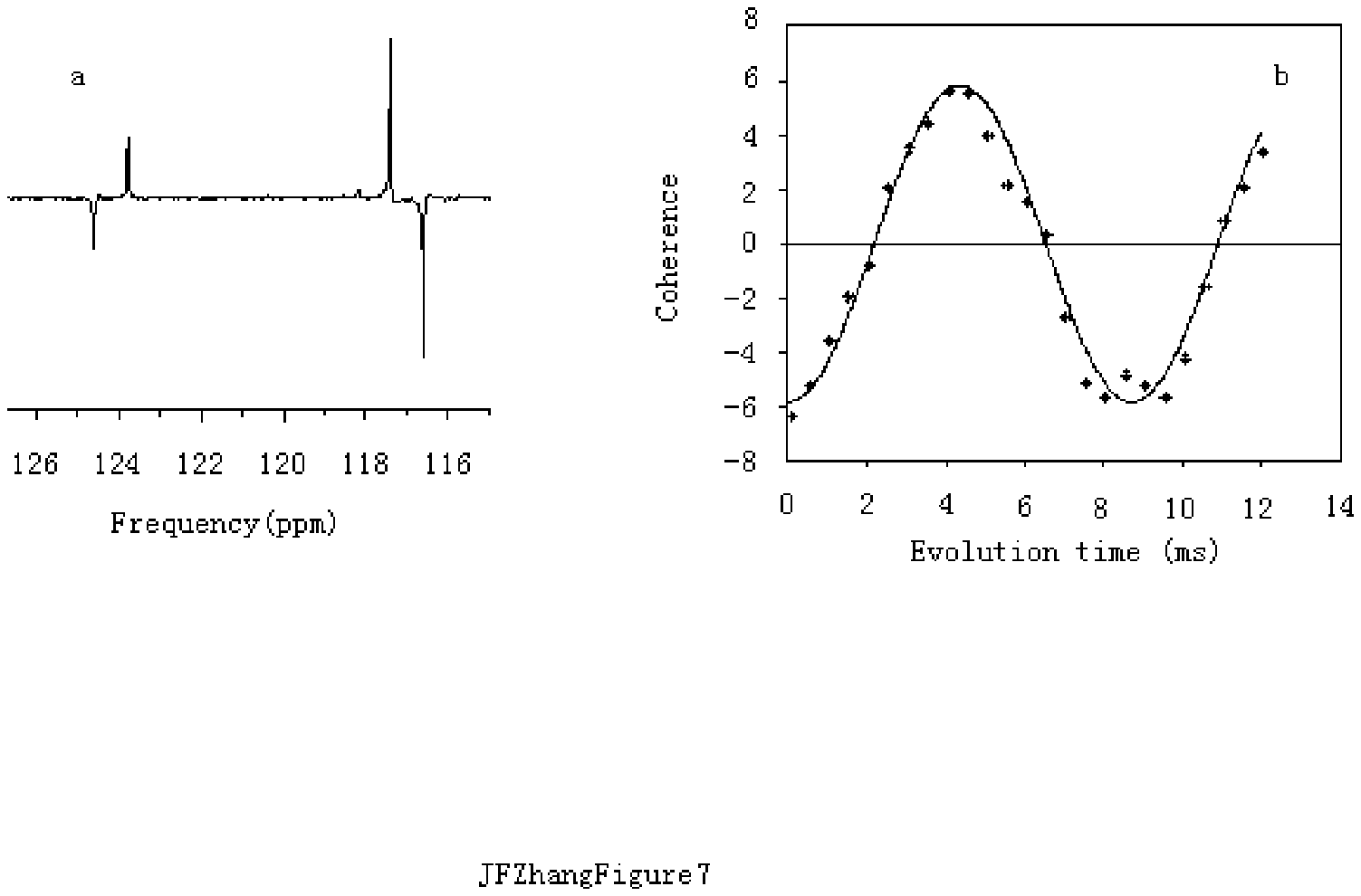}
\caption{}
\end{figure}
\end{document}